# Forecasting fuel combustion-related $CO_2$ emissions by a novel continuous fractional nonlinear grey Bernoulli model with Grey Wolf Optimizer

Wanli Xie[a], Wen-Ze Wu[b,*], Chong Liu[c], Tao Zhang[d], Zijie Dong[e]

[a]*Institute of EduInfo Science and Engineering, Nanjing Normal University, Nanjing 210097, China*

[b]*School of Economics and Business Administration, Central China Normal University, Wuhan 430079, China*

[c]*School of Science, Inner Mongolia Agricultural University, Hohhot 010018, China*

[d]*School of Science, Guangxi University of Science and Technology, Liuzhou 545006, China*

[e]*Faculty of Mathematics and Statistics, Hubei University, Wuhan 430062, China*

***corresponding author**. Wen-Ze Wu (wenzew@mails.ccnu.edu.cn)*

**Abstract:** Foresight of $CO_2$ emissions from fuel combustion is essential for policy-makers to identify ready targets for effective reduction plans and further to improve energy policies and plans. For the purpose of accurately forecasting the future development of China's $CO_2$ emissions from fuel combustion, a novel continuous fractional nonlinear grey Bernoulli model is developed in this paper. The fractional nonlinear grey Bernoulli model already in place is known that has a fixed first-order derivative that impairs the predictive performance to some extent. To address this problem, in the newly proposed model, a flexible variable is introduced into the order of derivative, freeing it from integer-order accumulation. In order to further improve the performance of the newly proposed model, a meta-heuristic algorithm, namely Grey Wolf Optimizer (GWO), is determined to the emerging coefficients. To demonstrate the effectiveness, two real examples and China's fuel combustion-related $CO_2$ emissions are used for model validation by comparing with other benchmark models, the results show the proposed model outperforms competitors. Thus, the future development trend of fuel combustion- related $CO_2$ emissions by 2023 are predicted, accounting for 10039.80 Million tons (Mt). In accordance with the forecasts, several suggestions are provided to curb carbon dioxide emissions.

# 1 Introduction

The topic focusing on carbon emission mitigation (Fang and Chen, 2019; Home-Ortiz et al., 2019) has been ongoing concern across countries on account of $CO_2$ being one of the most foremost greenhouse gases in the atmosphere that causes the global warming effect. Moreover, $CO_2$ from energy represents about 60% of the anthropogenic greenhouse gas emissions of global emissions. For purpose of controlling carbon emissions (Rocco et al., 2020), numerous efforts have been made by the international communities, such as the UNFCCC (United Nations Framework Convention on Climate Change), the Kyoto Protocol and the Paris Agreement. At the present stage, as the biggest emitter of $CO_2$, China has promulgated a continuously growing array of policies to curb carbon emissions. Particularly, the Chinese government has repeatedly stated that China is committed to reduce carbon intensity by 40-45% by 2020 and 60-65% by 2030, respectively, compared to the 2005 level. As a consequence, China promises to reach the peak of $CO_2$ by 2030. In the prevailing low-carbon economy, China is under huge pressure to reduce $CO_2$ emissions. As many studies reveal, $CO_2$ emissions from fuel combustion presents the largest share of the total. It is necessary to forecast fuel combustion-based $CO_2$ emissions, which could provide a solid basis for decision-makers to prepare and implement reasonable plans and policies.

In effect, there are diverse factors that have uncertain and complicated impact on final carbon emissions, making the forecast of $CO_2$ emissions from fuel combustion more arduous. To address this problem, many published papers have strived to predict $CO_2$ emissions by sector. For example, Köne and Büke (2010) made projections of $CO_2$ emissions from fuel combustion in selected countries by trend analysis. Pérez-Suárez and López-Menéndez (2015) combined environmental Kuznets curves with logistic growth models, so as to forecast $CO_2$ emissions for different countries. To forecasting province-level $CO_2$ emissions, Sun et al. (2017) put forward a novel PSO-ELM by incorporating factor analysis. Ding et al. (2020) estimated the future energy-related $CO_2$ emissions by a novel discrete grey multivariate model. From a conceptual perspective, these approaches is known as three categories - the statistical models, the machine learning methods and grey models.

As many literature suggest, statistical models have since been widely used in various disciplines due to its advantages of strong explanatory. For forecasting $CO_2$ emissions, Zhao and Du (2015) constructed an empirical panel regression which could be regarded as an extension of EKC literature, they projected the total emissions of sample countries in four scenarios according to country's emissions per capita. With help of multiple linear regression (MLR) and multiple polynomial regression (MPR) Hosseini et al. (2019) explored the carbon emissions by 2030 under assumptions of business-as-usual and the Sixth Development Plan. Ameyaw et al. (2019) designed a long short-term memory (LSTM) algorithm for forecasting $CO_2$ emissions from fossil fuel combustion only, this model was proven to be effective in the absence of exogenous variables and assumptions required. Malik et al. (2020) used the ARIMA model to forecast Pakistan's $CO_2$ emissions in the case of China-Pakistan Economic Corridor (CPEC) scenario. Belbute and Pereira (2020) employed an autoregressive fractionally integrated moving (ARFIMA) model to forecast fossil-fuel combustion and cement production-related $CO_2$ emissions for Portugal. It is evident that these models can obtain a satisfactory results and exhibit a desirable explanatory once the sufficient observations are given. However, the major limitation lies in that they require expert-knowledge to identify relationships between variables (Zhu et al., 2020), and the predictive accuracy heavily relies on parameter estimation.

As a type of advanced intelligent algorithms, machine learning methods could solve $CO_2$ emissions issues, due to its advantage of conducting complicated calculations. Qiao et al. (2020) considered neglected stability in the previous literature, and proposed a novel hybrid algorithm by combining genetic algorithm and lion swarm optimization to improve the performance of the traditional least squares, the results of forecasting carbon dioxide emissions of developed countries, showing the merits of stronger global optimization ability and higher accuracy. By taking the multiplex influential factors into considerations, Zhao et al. (2018) put forward a hybrid approach of Salp Swarm Algorithm (SSA) and Least Squares Support Vector Machine (LSSVM), further optimizing the model parameters by Particle Swarm Optimization (PSO), the results revealed this model enhanced the predictive capacity and reliability. Wen and Cao (2020) proposed a novel method for forecasting energy-related carbon emissions in Shanghai, they combined support vector machine and improved chicken swarm optimization embedded chaotic mutation and nonlinear weight index, and principal component analysis. In response to forecast the national carbon emissions in the building sector by region, Hong et al. (2018) studied an optimized gene expression programming model based on metaheuristic algorithms. For the sake of forecasting carbon emissions in commercial sector, Wen and Yuan (2020) used BP neural network on the basis of random forest and PSO. Other methods, including EBOA-LSSVM, PLS-Grey-Markov, CSCWOA-ELM, and so forth, ought to be referred to literature (Wen and Cao, 2020; Cui et al., 2018; Wang et al., 2018). The main drawback of machine learning methods is plenty of data that should be required. In contrast, if the sample size is small, the performance will be poor. In addition, the internal operation mechanism is unknown.

As introduced by Ofosu-Adarkwa et al. (2020), it is better to focus on the time sequence that involves the most relevant shock, attempting to modeling a time series impacted by shocks. In short, we should narrow the sequence closest to the current time to capture. Especially for China that is under economic restructuring and upgradation since China joined the World Trade Organization (WTO) in 2001. In this background, adequate information about annual carbon emissions by sector is unavailable. Grey forecasting model is known as an effective approach for forecasting the development trend of an uncertain system with sparse data. This is the essential motivation of this study to explore the future development trend of $CO_2$ emissions from fuel combustion by grey forecasting system.

In reality, there are many studies on forecasting carbon emissions by grey-based models. For example, Wu et al. (2015) established a multivariate grey model based on rolling mechanism to predict $CO_2$ emissions for the BRICS countries. Ding et al. (2017) reconstructed the traditional grey multivariate model by three steps -- optimizing the background value, adding the varying trend of driver variables and deriving the optimal time response - to forecast $CO_2$ emissions from fuel combustion of China. Wang and Li (2019) described the relationship between $CO_2$ emissions and economic growth in China through presenting a PSO-based grey Vehulst model. More recently, Ding et al. (2020) took nonlinear effects of the influential factors on emissions into considerations, further, Ding and his colleagues developed a novel discrete grey power model by plugging grey power indices into the model structure. In the same vein, Ofosu-Adarkwa et al. (2020) combined grey Vehulst model and GM(1,N) model, as a result, a hybrid model was proposed for forecasting $CO_2$ emissions from China's cement production. Wu et al. (2020) studied the properties of conformable fractional grey model and then applied this model to forecast $CO_2$ emissions in the BRICS countries. Admittedly, these models achieved satisfactory results as a whole. However, they are not without limitations in some certain situations. In consequence, this fact provide insights into further improving the predictive performance in forecasting $CO_2$ emissions from fuel combustion.

Inspired by the previous studies (see, *e.g.*, Ma et al., 2020; Wu et al., 2013), the fractional and conformable fractional grey models make grey forecasting theory fall from integer-order to fractional accumulation, which free it from integer-order accumulation. It is noticed that the order of derivative is still fixed as first order, which makes the derivative already in place discontinuous, thus impairing the predictive performance to some extent. With this in mind, in this paper a changing variable is introduced into the differential equation of the traditional conformable fractional nonlinear grey Bernoulli model (shorted for *CFNGBM(1,1)*), as a consequence, a novel grey model, namely continuous *CFNGBM(1,1)* (hereafter C*CFNGBM(1,1)* denoted as for short), is proposed. On the other hand, although the studies focusing on $CO_2$ emissions have increased over the last years, up-to-date and accurate projections merit further research, which provides a solid basis for decision-makers to prepare and frame reasonable plans and policies.

In view of claims mentioned above, aiming to attain more accurate forecasts of $CO_2$ emissions from fuel combustion, the major contributions of this paper are summarized as follows.

● Firstly, the fixed first order of derivative of *CFNGBM(1,1)* is replaced with a flexible and changing conformable fractional order. This model could provide decision-makers with the higher predictive accuracy of $CO_2$ emissions from fuel combustion compared with other benchmark models: grey-based and non-grey-based models.

● Secondly, the Grey Wolf Optimization (GWO) algorithm is applied to determine the emerging coefficients, namely fractional order of accumulation $r$, that of derivative $\alpha$ and nonlinear coefficient $\gamma$.

● Thirdly, The newly proposed model *CFNGBM(1,1)* is employed to model and predict $CO_2$ emissions from fuel combustion of China, in comparison with other benchmark models. The numerical results show the proposed model outperforms competitors. On this basis, the proposed model is more suitable for forecasting the future development trend of China's $CO_2$ emissions from fuel combustion from 2020 to 2023.

● Finally, based on the forecasts of this paper and the current situation of the Chinese economy, several suggestions are proposed for curbing $CO_2$ emissions in China.

The rest of this paper is organized as follows, Section 2 elaborates the modeling procedure of the newly proposed model. Section 3 verifies the effectiveness of the proposed model. Application of the newly proposed model to predict $CO_2$ emissions from fuel combustion is conducted in Section 4 and Section 5 concludes.

## 2 Methodology

As previously mentioned, the grey models have been successfully used in various fields. Owing to their superiority and applicability, the newly proposed model is employed to forecast China's fuel combustion-related $CO_2$ emissions.

### 2.1 Traditional *FNGBM(1,1)* model

The traditional *FNGBM(1,1)* model is committed to extend the *NGBM(1,1)* model (Chen, 2008) to more widely-used range by the fractional accumulation of original series (For more details Wu et al. (2019) should be referred). The modeling mechanism can be summarized as follows.

**Step one.** Obtain the original series and its *r*-order fractional accumulative generating operation sequence. Let the original series be $X^{(0)} = \left(x^{(0)}(1), x^{(0)}(2), \cdots, x^{(0)}(n)\right)$, and then the *r*-order fractional accumulative generating operation sequence is given by $X^{(1)} = A^r X^{(0)} = \left(x^{(1)}(1), x^{(1)}(2), \cdots, x^{(1)}(n)\right)$, where

$$A^r = \begin{pmatrix} \begin{bmatrix} r \\ 0 \end{bmatrix} & 0 & 0 & \cdots & 0 \\ \begin{bmatrix} r \\ 1 \end{bmatrix} & \begin{bmatrix} r \\ 0 \end{bmatrix} & 0 & \cdots & 0 \\ \begin{bmatrix} r \\ 2 \end{bmatrix} & \begin{bmatrix} r \\ 1 \end{bmatrix} & \begin{bmatrix} r \\ 0 \end{bmatrix} & \cdots & 0 \\ \vdots & \vdots & \vdots & \ddots & \vdots \\ \begin{bmatrix} r \\ n-1 \end{bmatrix} & \begin{bmatrix} r \\ n-2 \end{bmatrix} & \begin{bmatrix} r \\ n-3 \end{bmatrix} & \cdots & \begin{bmatrix} r \\ 0 \end{bmatrix} \end{pmatrix}$$

with $\begin{bmatrix} r \\ i \end{bmatrix} = \dfrac{r(r+1)\cdots(r+i-1)}{i!} = \dfrac{(r+i-1)!}{i!(r-1)!} = \begin{pmatrix} r+i-1 \\ i \end{pmatrix}$, $\begin{bmatrix} 0 \\ i \end{bmatrix} = 0$, $\begin{bmatrix} 0 \\ 0 \end{bmatrix} = \begin{pmatrix} 0 \\ 0 \end{pmatrix} = 1$.

**Step two.** Define the mathematical formula of *FNGBM(1,1)*. With reference to Wu et al. (2019), the differential equation of *FNGBM(1,1)* is argued as

$$\frac{dx^{(r)}}{dt} + ax^{(r)} = b\left(x^{(r)}\right)^{\gamma} \quad (1)$$

and the discrete form of Eq. (1) can be obtained as

$$x^{(r)}(k) - x^{(r)}(k-1) + az^{(r)}(k) = b\left(z^{(r)}(k)\right)^{\gamma} \quad (2)$$

where the background value $z^{(r)}(k) = 0.5\left(x^{(r)}(k) + x^{(r)}(k-1)\right)$.

**Step three.** Estimate the structure parameter vector $\psi = (a,b)^{\mathrm{T}}$. Through the least square method, the structure parameter vector $\psi = (a,b)^{\mathrm{T}}$ can be obtained:

$$\psi = (a,b)^{\mathrm{T}} = \left(\mathbf{B}^{\mathrm{T}}\mathbf{B}\right)^{-1}\mathbf{B}^{\mathrm{T}}\mathbf{Y} \quad (3)$$

where

$$\mathbf{B} = \begin{pmatrix} -z^{(r)}(2) & \left(z^{(r)}(2)\right)^{\gamma} \\ -z^{(r)}(3) & \left(z^{(r)}(3)\right)^{\gamma} \\ \vdots & \vdots \\ -z^{(r)}(n) & \left(z^{(r)}(n)\right)^{\gamma} \end{pmatrix}, \quad \mathbf{Y} = \begin{pmatrix} x^{(r)}(2) - x^{(r)}(1) \\ x^{(r)}(3) - x^{(r)}(2) \\ \vdots \\ x^{(r)}(n) - x^{(r)}(n-1) \end{pmatrix}$$

**Step four.** Calculate the time function for *FNGBM(1,1)*. Assume that $x^{(r)}(1) = x^{(0)}(1)$, then the time function of *FNGBM(1,1)* is suggested as

$$\hat{x}^{(r)}(k) = \left\{ \left[ \left(x^{(0)}(1)\right)^{1-\gamma} - \frac{b}{a} \right] e^{-a(1-\gamma)(k-1)} + \frac{b}{a} \right\}^{\frac{1}{1-\gamma}} \quad (4)$$

**Step five.** Acquire restored values from simulative values. In order to obtain the predicted values, the r-order inverse fractional accumulative generating operation is introduced as

$$\hat{x}^{(0)}(k) = \hat{x}^{(r)}(k) - \hat{x}^{(r)}(k-1), k = 2, 3, \cdots \quad (5)$$

**2.2 The newly proposed *CCFNGBM(1,1)* model**

As stated by the previous literature, admittedly, the fractional accumulative operations improve the performance of *NGBM(1,1)*, while just like other grey models, the differential equations of this model is still integer-order, which also impairs the accuracy of the model to some extent. To this end, conformable fractional derivative is introduced, further improving the prediction precision and generalization of *FNGBM(1,1)*. The modeling procedure of *CCFNGBM(1,1)* is studied in the coming subsections.

*2.2.1 Presentation of conformable fractional accumulation and difference*

Inspired by Khalil and Abu-Shaab (2015), Ma et al. (2020) propose the definition of conformable fractional accumulation and its inverse operator, which will be in lieu of integer-order derivative in the current study.

**Definition 1.** Given a differential function $f:[0,\infty) \to R$, the conformable fractional accumulation (*CFA*) of $f$ with $\alpha$ order is defined as

$$\begin{cases} \nabla^\alpha f(k) = \nabla\left(\dfrac{f(k)}{k^{1-\alpha}}\right) = \sum_{j=1}^{k}\left(\dfrac{f(k)}{j^{1-\alpha}}\right) & k \in N^+, \alpha \in (0,1] \\ \nabla^\alpha f(k) = \nabla^n\left(\dfrac{f(k)}{k^{[\alpha]-\alpha}}\right) & k \in N^+, \alpha \in (n, n+1], n \in N \end{cases} \quad (6)$$

where $[\cdot]$ is the ceil function, *i.e.* the $[\alpha]$ is the smallest integer no larger than $\alpha$. In the case of no loss of generality, $\nabla^\alpha$ denotes the conformable fractional accumulation.

**Definition 2.** The conformable fractional difference (*CFD*) of $f$ with $\alpha$ order is defined as

$$\begin{cases} \Delta^\alpha f(k) = k^{1-\alpha}\Delta f(k) & \alpha \in (0,1], k \in N^+ \\ \Delta^\alpha f(k) = k^{[\alpha]-\alpha}\Delta^{n+1} f(k) & \alpha \in (n, n+1], k \in N^+, n \in N \end{cases} \quad (7)$$

*2.2.2 Building the CCFNGBM(1,1) model*

Assume that $X^{(0)}, X^{(r)}$ are mentioned in Section 2.1, the differential equation of *CCFNGBM(1,1)* is expressed by

$$\dfrac{d^\alpha x^{(r)}}{dt^\alpha} + ax^{(r)} = b\left(x^{(r)}\right)^\gamma, r \in R^+, \alpha \in (0,1], \gamma \in R \quad (8)$$

where $a$ and $b\left(x^{(r)}\right)^\gamma$ denote the development coefficient and grey action quantity; $r, \alpha, \gamma$ are the fractional order, conformable fractional order and nonlinear parameter, respectively.

**Remark 1.** *As seen in Eq. (8), when $\alpha = 1$ and $r = 1$, CCFNGBM(1,1) reduces to NGBM(1,1); when $a = r = 1$ and $\gamma = 2$, CCFNGBM(1,1) equals to grey Vehulst model.*

**Theorem 1.** The exact solution to the *CCFNGBM(1,1)* model is given by

$$\hat{x}^{(r)}(k) = \left\{ \left[ \left(x^{(0)}(1)\right)^{1-\gamma} - \frac{b}{a} \right] e^{-\frac{a}{\alpha}(1-\gamma)(k^\alpha - 1)} + \frac{b}{a} \right\}^{\frac{1}{1-\gamma}} \quad (9)$$

**Proof.** See Appendix A.

Subsequently, in order to estimate the structural parameter vector $\boldsymbol{\beta} = (a,b)^{\mathrm{T}}$, the discrete formula of Eq. (8) should be derived, as previously stated. Firstly, integrating the both sides of Eq. (8), which yields

$$\int\!\!\cdots\!\!\int_{k-1}^{k} \frac{d^\alpha x^{(r)}}{dt^\alpha} dt^\alpha + a \int\!\!\cdots\!\!\int_{k-1}^{k} x^{(r)} dt^\alpha = b \int\!\!\cdots\!\!\int_{k-1}^{k} \left(x^{(r)}\right)^\gamma dt^\alpha \quad (10)$$

In accordance with trapezoid approximation formula, we have

$$\begin{cases} \int\!\!\cdots\!\!\int_{k-1}^{k} \frac{d^\alpha x^{(r)}}{dt^\alpha} dt^\alpha = \nabla^\alpha x^{(r)} = x^{(r-\alpha)}(k) \\ \int\!\!\cdots\!\!\int_{k-1}^{k} x^{(r)} dt^\alpha \approx \frac{x^{(r)}(k) + x^{(r)}(k-1)}{2} = z^{(r)}(k) \\ \int\!\!\cdots\!\!\int_{k-1}^{k} \left(x^{(r)}\right)^\gamma dt^\alpha \approx \left(\frac{x^{(r)}(k) + x^{(r)}(k-1)}{2}\right)^\gamma = \left(z^{(r)}(k)\right)^\gamma \end{cases} \quad (11)$$

Substituting Eq. (11) into Eq. (10), one can write

$$x^{(r-\alpha)}(k) + a z^{(r)}(k) = b \left(z^{(r)}(k)\right)^\gamma \quad (12)$$

If we let

$$\mathbf{B} = \begin{pmatrix} -z^{(r)}(2) & \left(z^{(r)}(2)\right)^\gamma \\ -z^{(r)}(3) & \left(z^{(r)}(3)\right)^\gamma \\ \vdots & \vdots \\ -z^{(r)}(n) & \left(z^{(r)}(n)\right)^\gamma \end{pmatrix}, \quad \mathbf{Y} = \begin{pmatrix} x^{(r-\alpha)}(2) \\ x^{(r-\alpha)}(3) \\ \vdots \\ x^{(r-\alpha)}(n) \end{pmatrix}$$

then the structure parameter vector $\boldsymbol{\beta}$ can be obtained using the least square method, which is

$$\boldsymbol{\beta} = (a,b)^{\mathrm{T}} = (\mathbf{B}^{\mathrm{T}}\mathbf{B})^{-1}\mathbf{B}^{\mathrm{T}}\mathbf{Y} \quad (13)$$

Substituting Eq. (13) into Eq. (9), the predicted values of $x^{(r)}(k), k=2,3,\cdots$ can be acquired, eventually achieving the forecast for $CO_2$ emissions from fuel combustion, with the help of the inverse calculation shown in Eq. (5). The newly proposed model not only prefects the family of grey forecasting models but also improves the existing studies that focus on fractional grey models.

**2.3 Searching the emerging coefficients for *CCFNGBM(1,1)***

As seen in Section 2.2, the emerging coefficients, namely the conformable fractional order $r$, the continuous differential order $\alpha$ and the nonlinear coefficient $\gamma$, are unknown and need to be estimated, it is a crucial step in the modelling procedure of the newly proposed model due to the quality of the emerging coefficients having a great impact on the final prediction performance.

For the sake of searching the optimal coefficients for the newly proposed model, as repeatedly stated in several relevant papers (see, *e.g.*, Wang et al., 2019; Ding et al., 2020; Zhou et al., 2020). There is a simple optimization problem with constraints designed for searching the optimal coefficients, the mathematical formula can be defined as

$$\arg\min f(r,\alpha,\gamma) = \frac{1}{n-1}\sum_{k=2}^{n} \frac{\left|\hat{x}^{(0)}(k) - x^{(0)}(k)\right|}{x^{(0)}(k)} \times 100\% \quad (14)$$

$$s.t. \begin{cases} \boldsymbol{\beta} = (a,b)^{\mathrm{T}} = (\mathbf{B}^{\mathrm{T}}\mathbf{B})^{-1}\mathbf{B}^{\mathrm{T}}\mathbf{Y} \\ \hat{x}^{(r)}(k) = \left\{ \left[ (x^{(0)}(1))^{1-\gamma} - \frac{b}{a} \right] e^{-\frac{a}{\alpha}(1-\gamma)(k^{\alpha}-1)} + \frac{b}{a} \right\}^{\frac{1}{1-\gamma}} \end{cases}$$

where $x^{(0)}(k), \hat{x}^{(0)}(k)$ represent the actual value and its corresponding predictive value at point $k$.

The above equation appears to be difficultly solved through an ordinary method due to it having nonlinear characteristics. To address this problem, a meta-heuristic algorithm, namely Grey Wolf Optimizer (GWO), is selected to search the optimal emerging coefficients in the current study.

GWO, proposed by Mirjalili et al. (2014), stems from the imitation of the social hierarchy and hunting behaviors of the grey wolves. In GWO, the wolves are labeled into four levels, namely $\alpha, \beta, \delta, \omega$ according to their social hierarchies, where $\alpha$ denotes the fittest solution (or, best candidate agent), followed by $\beta, \delta, \omega$. Generally speaking, the hunting is guided by $\alpha, \beta, \delta$. The distance of the wolves is expressed as

$$\mathbf{D} = |\mathbf{C}\mathbf{X}_p(t) - \mathbf{X}(t)| \quad (15)$$

$$\mathbf{X}(t+1) = \mathbf{X}_p(t) - \mathbf{A} \cdot \mathbf{D} \quad (16)$$

where $\mathbf{X}_p$ refers to the position of the prey, and $\mathbf{X}$ denotes the position of a grey wolf; $\mathbf{A}, \mathbf{C}$ are coefficient vectors calculated as

$$\mathbf{A} = 2 \cdot \mathbf{a} \cdot \mathbf{r}_1 - \mathbf{a} \quad (17)$$

$$\mathbf{C} = 2\mathbf{r}_2 \quad (18)$$

where components of $\mathbf{a}$ are linearly declined from 2 to 0 for each iteration, both $\mathbf{r}_1, \mathbf{r}_2$ are random vectors generated within the interval $[0,1]$.

After saving the first three best solutions obtained so far, then obliging other search agents to update their positions in accordance with the best search agent, the mathematical relevant formula is given as

$$\mathbf{X}(t+1) = \frac{\mathbf{X}(t)}{3} \quad (19)$$

For more details Mirjalili et al. (2014) and Ma et al. (2019) should be referred, the pseudo-code of the GWO algorithm-based *CCFNGBM(1,1)* model can be seen in Algorithm 1.

**2.4 Model evaluation measure**

To assess the degree of the prediction accuracy of the newly proposed model, there are two commonly-used indicators, namely Absolute Percentage Error (*APE*) and Mean Absolute Percentage Error (*MAPE*), selected as the judgement methods to evaluate the newly proposed model, which are defined as follows.

$$APE = \frac{\left|\hat{x}^{(0)}(k) - x^{(0)}(k)\right|}{x^{(0)}(k)} \cdot 100\% \quad (20)$$

$$MAPE = \frac{1}{n-1} \sum_{k=2}^{n} \frac{\left|\hat{x}^{(0)}(k) - x^{(0)}(k)\right|}{x^{(0)}(k)} \cdot 100\% \quad (21)$$

## 3 Validation of the newly proposed model

In this section, there are two numerical examples taken to testify the validity of the newly proposed model through comparing to other grey forecasting models, including *GM(1,1)* (Deng, 1982), *DGM(1,1)* (Xie and Liu, 2009), *NGBM(1,1)* and *FNGBM(1,1)*. Notice that there are unknown coefficients of *NGBM(1,1)* and *FNGBM(1,1)* are also determined by GWO to ensure that the comparative analysis of this study is set at the same level.

***Case 1.*** (Forecasting diesel fuel consumption of Shanghai [unit: $10^4$t]) In this case, diesel fuel consumption of Shanghai from 2000 to 2017 are taken as an example to verify the validity of the proposed model, which are collected from the National

Bureau of Statistics of China. The data starting from 2000 to 2015 are chosen to calibrate the model and the left two samples are used to examine the accuracy. To be specific, in *CCFNGBM(1,1)* the emerging coefficients are obtained as $r = 0.6660, \alpha = 0.0958, \gamma = 6.0956$, with the help of GWO. Fig. 1 vividly describes the track of seeking the optimal coefficients by GWO. Moreover, as seen in Fig. 2, over the course of the latest ten sub-generations, the search target is increasingly focusing within a small varying range. This fact shows GWO effectively and steadily determines the emerging coefficient for the newly proposed model. With reference to the existing relevant references, predictive results of the competitive models can be obtained as shown in Table 1. It is evident that the predictive performance produced by the newly proposed model is best either in the simulative stage or in the predictive stage, indicating the proposed model outperforms competitors in this case.

*Case 2.* (Forecasting total CO2 emissions of Germany [unit: $10^4$t]) In this case, total CO2 emissions of Germany from 2008 to 2018 are employed to validate the predictive capacity of the newly proposed model, which are collected from the *BP Statistical Review of World Energy 2019*. Similar to Case 1, we categorize the raw data into two groups, the first 8 samples are used to establish the predictive model and the left two samples are used to detect the accuracy. As Fig. 3 shows, with the help of GWO, the optimal emerging coefficients are $r = 1.6713, \alpha = 0.3242, \gamma = 0.9512$ respectively. Fig. 4 proves the robustness of GWO in the process of searching the emerging coefficients (elaborated in Case 1). All predictive results by five competitive models are shown in Table 2. As for the simulative stage, although the MAPE value is not best, it is very close to the best value produced by NGBM(1,1). In particular, the MAPE is much smaller than that of the sub-optimal model in the predictive stage. This fact concludes that the proposed model has a better performance in this case.

## 4 Application

In view of the superiority of the newly proposed model introduced in Section 3, it is applied to make a projection of $CO_2$ emissions from fuel combustion of China. This section is divided into four subsections, including data description, experimental design, comparative analysis and future emissions forecasting. The flowchart of forecasting fuel combustion-related $CO_2$ emissions by the newly proposed model is graphed in Fig. 5.

### 4.1 Data description

We consider annual data for $CO_2$ emissions from fuel combustion in China for the period between 2001 and 2019. This is driven by a fact that Chinese economy and $CO_2$ emissions increased slowly before 2000, while they grew quickly after 2000, an important reason is that China joining the World Trade Organization (WTO) in 2001, entering a market economy country instead of a planned economy country (Ding et al., 2020). Information before 2001 has a little reference value according to the principle of priority of new information. In addition, Ofosu-Adarkwa et al. (2020) pointed that in modeling a sequence that was impacted by shocks, it was better to focus on the most relevant shocks, narrowing the series closet to the current time to capture. As such, the above data sequence is powered in making future projections of $CO_2$ emissions from fuel combustion.

We obtained emissions between 2001 and 2019 by the Statistical Review of World Energy, in which, to be specific, the carbon emissions reflect only those through consumption of oil, gas and coal for combustion related activities. In addition, we take $CO_2$ emissions from fuel combustion reported by the International Energy Agency (IEA) into considerations, which is divided into four components - coal, oil, gas, international marine bunkers, international aviation bunkers, $CO_2$ emissions from fuel combustion by sectors are also recorded. As Fig. 6 argues, allowing for statistical bias, the two series are recognized basically consistent for the period they overlap with quest to make future projections in this thesis.

Fig. 6 exhibits that $CO_2$ emissions from fuel combustion of China has experienced extensive increase, accounting for 3593.1 Mt in 2001 however 9920.5 Mt in 2019, there is over 2.5 times the original. With reference to Fig. 7, it is evident that there are obvious discrepancies between different sectors, for example, electricity and heat production occupy the largest share by 50%, followed by manufacturing industries and construction, transport. The sum of other sectors (*e.g.* residential, commercial and public services, other energy industry own use, among others) arrives at 10%. $CO_2$ emissions from fuel combustion of China is closely related to energy structure, observing China's $CO_2$ emissions from fuel combustion from an angel of energy structure, as shown in Fig. 8, it is found that coal combustion is a dominated driver contributing to China's $CO_2$ emissions from fuel combustion, arriving at 2689.1 Mt in 2001 instead 7469.9 Mt in 2017, with a share of about 80% of total emissions.

## 4.2 Model calibration for the newly proposed model

For the purpose of demonstrating the superiority of the newly proposed model, apart from the grey-based models (specified in Section 3), the non-grey-based benchmark models, namely statistical models and machine learning methods, are also established in the current work. The PR (Polynomial Regression) (Akhlaghi et al. 2019) and ARIMA (Autoregressive integrated Moving Average) (Singh et al., 2019) is among the building blocks of statistical models. The ANN (Artificial Neural Network) (Afram et al., 2017) and SVM (Support Vector Machine) (Richhariya and Tanveer, 2018) be conceived as the representative of machine learning methods. By the way of comparing with these competitive models, the effectiveness and superiority of the newly proposed model can be verified comprehensively. In particular, the data from 2001 to 2017 are used for model calibration and the left two observations are chosen to testify the accuracy. Therefore, the modeling steps of the newly proposed model based on the above introduction can be summarized as follows.

**Step one.** Obtain the original series $X^{(0)} = (3593.1, 3910.6, \cdots, 9396.9)$, both the r-order conformable fractional accumulation sequence $X^{(r)}$ and background value $Z^{(r)}$ can be acquired when initializing $r$ to 1.

**Step two.** Initialize $\alpha, \gamma$ to 1 and, with the aid of the least square method, the model parameter vector can be calculated by $\boldsymbol{\beta} = (a,b)^{\mathrm{T}} = (\mathbf{B}^{\mathrm{T}}\mathbf{B})^{-1}\mathbf{B}^{\mathrm{T}}\mathbf{Y}$, which yields that $\begin{cases} a = -0.0439 \\ b = 5136.16 \end{cases}$.

**Step three.** Calculate the time response function:

$$\hat{x}^{(r)}(k) = \left\{\left(3593.10^{1-\gamma} + 117030.89\right)e^{0.0439(1-\gamma)(k-1)} - 117030.89\right\}^{\frac{1}{1-\gamma}}$$

**Step four.** Optimizing the model parameter vector **β** by taking different values of the emerging coefficients, with the following fitness function,

$$fitness(r,\alpha,\gamma) = \frac{1}{n-1}\sum \frac{\left|\alpha^{(-r)}\left\{\left(3593.10^{1-\gamma} + 117030.89\right)e^{0.0439(1-\gamma)(k-1)} - 117030.89\right\}^{\frac{1}{1-\gamma}} - x^{(0)}(k)\right|}{x^{(0)}(k)} 100\%$$

With help of the GWO algorithm, the optimal emerging coefficients $r, \alpha, \gamma$ that correspond to the best fitness value can be searched as [0.938, 0.3037, 1.3164]. In the meantime, the track of seeking process by GWO is shown in Fig. 9.

**Step five.** Substituting the model parameters into Steps 1-3, the time response function of the newly proposed model is given by

$$\hat{x}^{(r)}(k) = \left(0.0613 e^{-0.3686(k^{0.3037}-1)} + 0.0137\right)^{-3.1605}, k=2,3,\cdots,n,n+1,\cdots$$

**4.3 Comparative analysis**

In view of the implementation of the above procedure, the predictive results by the newly proposed model and eight benchmark models can be calculated, as shown in Tables 3 and 4.

The APEs of the newly proposed model and *FNGBM(1,1)* vary in a small range, [0%, 5%] at the entire sequence, while those of others enjoy wider fluctuations with at least [0%, 6%], indicating the fractional order accumulation is considered as an effective approach to improve the prediction performance significantly. To be specific, the range of APEs with [0.03%, 2.54%] by the newly proposed model is narrower for the predictive stage, this fact allows us to believe that the newly proposed model outperforms competitive models.

For MAPE, their values by these nine models are less than 10%, meaning they are appropriate for modeling $CO_2$ emissions from fuel combustion according to Lewis standard (Lewis, 1982). Nevertheless, the newly proposed model enjoys the lowest MAPE value of 1.33% for the predictive stage, suggesting that this model has more outstanding ability to forecast $CO_2$ emissions from fuel combustion compared to other benchmark models. Analyzing the modeling procedure of these models, the *NGBM(1,1)* capture nonlinear characteristics, so as to obtain a more desirable outcome relative to linear models, namely *GM(1,1)* and *DGM(1,1)*. In addition, the coupled combination of fractional accumulation and differential make the *NGBM(1,1)* model better. As a machine learning method that requires large sample size, the ANN and *SVM* do not acquire the best results, so are the statistical models (*i.e. ARIMA(2,1,0)* and *PR(2)*), this is primarily due to the small data sequence.

Figs. 10 and 11 are committed to visibly display line graph and error distribution by these competitors. They support the above findings that the newly proposed model can match well the predictive values with the observations, and produce lower predictive derivation from a graphic perspective.

**4.4 Fuel combustion-based $CO_2$ emissions forecasts**

In the wake of investigating the superiority of the newly proposed model, there is the pressing need to apply this model to forecast the future emissions from fuel combustion, which could provide a basis for decision-makers to prepare the suitable plans and policies.

Driven by the CCFNGBM(1,1) model as the above subsection introduces, $CO_2$

emissions from fuel combustion starting from 2020 to 2023, which are drawn in Fig. 12, there is an upward trend by 2023, accounting for 9860.35 Mt in 2020 and 10039.80 Mt in 2023, supporting the findings in Ding et al. (2020) that energy-related $CO_2$ emissions may reach 9936 Mt. Our results are relatively reasonable due to referring to the official dataset that are reported by International Energy Agency (IEA, 2017) (10205 Mt) and Economic Forecasting Division of the State Information Center (EFDSI, 2016) (9860 Mt), respectively. In addition, the growth rate of fuel combustion-based $CO_2$ emissions will be experienced a downward trend in the future years, basically below 1%, this increasing trend is aligned with the target promised by the Chinese government, having a potential to achieve the target of peaking $CO_2$ emissions by 2030 or earlier.

## 5 Conclusion and suggestions

Fuel combustion-based carbon emissions forecasting is crucial for framing and implementing the reasonable plans and policies, this is primarily due to diverse national energy structures. To accurately make projections of $CO_2$ emissions from fuel combustion of China by 2023, this paper develops a novel continuous fractional *NGBM(1,1)* model, namely *CCFNGBM(1,1)*, by simultaneously incorporating conformable fractional accumulation and derivative into the traditional *NGBM(1,1)* model that is capable of capturing the nonlinear characteristics hidden in sequences. To further improve the predictive capacity of the newly proposed model, this study traps GWO (Grey Wolf Optimizer) into determining the emerging coefficients. This model can not only perfect the grey forecasting model falling from integer-order to fractional derivative but also provide relatively reliable forecasts for decision-makers.

Two examples already in place are employed to verify the superiority of the newly proposed model comparing with other grey-based models by the introduce of the MAPE (Mean Absolute Percentage Error). After that, this model is applied to model and predict $CO_2$ emissions from fuel combustion of China, in comparison with other benchmark models, including grey-based and non-grey-based models, the predictive capacity of the newly proposed model is testified again. The future value in 2023 is expected to reach 10039.80 Mt, while the growth rate fall significantly, which could provide a solid basis for decision-maker to prepare the following plans and polices.

Based on the analysis in Section 4, combined the forecasts by this thesis with the current situation of China, several suggestions on curb $CO_2$ emissions from fuel combustion are presented as follows.

● Developing low-carbon technologies. As Fig. 6 suggests, the largest share of $CO_2$ emissions from fuel combustion is electricity and heat production, this process produces about 50% of total carbon emissions of China. Accelerating the progress of low-carbon technology is essential for green economic growth of China. Low-carbon technologies involve power, transportation, construction, metallurgy, chemical, petrochemical and other sectors, as well as developments in renewable energy and new energy, clean and efficient use of coal, exploration and development of oil and gas resources and coalbed methane, and carbon dioxide capture and storage. New technologies to effectively control greenhouse gas emissions.

● Accelerating the promotion of the national carbon market. Apart from the electricity and heat production, the manufacturing industry is also a consumer of high energy consumption, making economic restructuring and industrial upgrading a difficult task. Coal consumption still accounts for more than 50% of China's energy production, and the intensity of carbon dioxide emissions per energy unit is 30% higher than the world average. Energy consumption per unit of GDP is still very high, about 1.5 times the world average and 2 to 3 times that of developed countries. framing an action plan for peak carbon dioxide emissions and accelerating the promotion of the national carbon market. China must make greater efforts than developed countries to achieve carbon neutrality.

● Strengthening citizens' environmental awareness. On the basis of establishing citizens' environmental rights, further expanding their vision of environmental ethics and raising awareness of environmental issues. We must mobilize all aspects of society to give full play to the role of organizations at all levels, including agencies, media, schools, units, and communities, to reach the masses. Vigorously publicizing the significance of environmental protection, make people aware of the hazards of environmental problems, and enhance the sense of responsibility and mission of environmental protection.

Of course the proposed method is not without limitations, for example, the results we forecast by using the newly proposed model are conservative, this is primarily due to the sparse historical records we take into considerations with the mechanism of grey forecasting models. On the other hand, the ease of implementation is at cost of ignoring the potential factors related to fuel combustion-based carbon emissions, such as the level of urban, that of industrialization, energy structure, and so forth, these issues need to be addressed in the following research.

**Funding Information**

This work was supported by the Fundamental Research Funds for the Central Universities of China (Grant no. 2019YBZZ062), the Postgraduate Research & Practice Innovation Program of Jiangsu Province (Grant no. KYCX20_1144), the National Natural Science Foundation of China (Grant no. 11861014) and the Natural Science Foundation of Guangxi (Grant no. 2018GXNSFAA281145).

**Appendix A.** Proof of Theorem 1.

**Proof.** Multiplying the both sides of Eq .(8) by $\left(x^{(r)}(t)\right)^{\gamma}$, we have

$$\frac{d^{\alpha} x^{(r)}(t)}{dt^{\alpha}}\left(x^{(r)}(t)\right)^{\gamma} + a\left(x^{(r)}(t)\right)^{1-\gamma} = b \quad (A.1)$$

Setting $y^{(r)}(t) = \left(x^{(r)}(t)\right)^{1-\gamma}$, Eq. (A.1) can be rewritten as

$$\frac{d^{\alpha} y^{(r)}(t)}{dt^{\alpha}} + a(1-\gamma) y^{(r)}(t) = b(1-\gamma) \quad (A.2)$$

According to the property of conformable fractional derivative (Khalil, 2014), $T_{\alpha}(f)(t) = t^{1-\alpha} \frac{df(t)}{dt}$, noting that $T_{\alpha}(f)(t)$ and $\frac{df(t)}{dt}$ represent $\alpha-$ order conformable fractional derivative and integer-order derivative in Riemannian Geometry, respectively. Eq. (A.2) becomes

$$\left(y^{(r)}(t)\right)^{1-\alpha} \frac{dy^{(r)}(t)}{dt} + a(1-\gamma) y^{(r)}(t) = b(1-\gamma) \quad (A.3)$$

Solving Eq. (A.3) yields that

$$y^{(r)}(t) = e^{-\int(1-\gamma)\frac{a}{t^{1-\alpha}}dt}\left[\int(1-\gamma)\frac{b}{t^{1-\alpha}}e^{-\int(1-\gamma)\frac{a}{t^{1-\alpha}}dt}dt + C\right] \quad (A.4)$$

Further,

$$x^{(r)}(t) = \left\{e^{-\int(1-\gamma)\frac{a}{t^{1-\alpha}}dt}\left[\int(1-\gamma)\frac{b}{t^{1-\alpha}}e^{-\int(1-\gamma)\frac{a}{t^{1-\alpha}}dt}dt + C\right]\right\}^{\frac{1}{1-\gamma}} \quad (A.5)$$

Assume that $x^{(r)}(1) = x^{(0)}(1)$, we obtain

$$C = \left[\left(x^{(0)}(1)\right)^{1-\gamma} - \frac{b}{a}\right]e^{a\frac{1}{\alpha}(1-\gamma)} \quad (A.6)$$

Substituting Eq. (A.6) into Eq. (A.5), we have

$$x^{(r)}(t) = \left\{\left[\left(x^{(0)}(1)\right)^{1-\gamma} - \frac{b}{a}\right]e^{-a\frac{1}{\alpha}(1-\gamma)(t^{\alpha}-1)} + \frac{b}{a}\right\}^{\frac{1}{1-\gamma}} \quad (A.7)$$

Setting $t = k$, this completes the proof.

# Figure list

**Fig. 1** The track of seeking the emerging coefficients by GWO.

**Fig. 2** The process of seeking the emerging coefficients during the latest ten iterations based on GWO.

**Fig. 3** The track of seeking the emerging coefficients by GWO.

**Fig. 4** The process of seeking the emerging coefficients during the latest ten iterations based on GWO.

**Fig. 5** Structural flowchart of the newly proposed model in forecasting $CO_2$ emissions from fuel combustion of China.

**Fig. 6** Statistical data of $CO_2$ emissions from fuel combustion of China.

**Fig. 7** $CO_2$ emissions from fuel combustion by sector.

**Fig. 8** Development trend of China's $CO_2$ emissions from different fuel combustion.

**Fig. 9** The track of seeking process by GWO.

**Fig. 10** Curves of raw data and predictive results by nine competitors.

**Fig. 11** Error distribution of nine comparative models

**Fig. 12** The future development trend of $CO_2$ emissions from fuel combustion of China.

**Algorithm 1: The algorithm of GWO to seek the optimal coefficients** $\varphi = (r, \alpha, \gamma)$

Set the fitness function and initialize the parameters for GWO

Input: The original series $x^{(0)}$

Output: The optimal coefficients $\varphi = (r, \alpha, \gamma)$

for $t < T$ do

   Substitute $\varphi = (r, \alpha, \gamma)$ into Eq. (13) to obtain parameter vector $\hat{\beta} = (a, b)^{\mathrm{T}}$

   Substitute parameter vector into Eq. (9) to calculate the simulative values $\hat{x}^{(r)}$

   Calculate the restored values $\hat{x}^{(0)}$ in Eq. (5)

   Compute the MAPE value in Eq. (14)

end

   Update the minimum value MAPE

Return the best coefficients $\varphi = (r, \alpha, \gamma)$

Table 1. The results by five grey models in forecasting diesel fuel consumption of Shanghai (unit: $10^4$ t)

| Year | Data | GM | DGM | NGBM | FNGBM | CCFNGBM |
|---|---|---|---|---|---|---|
| 2000 | 176.44 | 176.44 | 176.44 | 176.44 | 176.44 | 176.44 |
| 2001 | 231.80 | 276.89 | 277.48 | 196.26 | 205.12 | 210.14 |
| 2002 | 236.78 | 293.53 | 294.10 | 245.81 | 236.78 | 251.08 |
| 2003 | 288.32 | 311.16 | 311.72 | 288.32 | 270.97 | 288.31 |
| 2004 | 346.56 | 329.86 | 330.39 | 325.89 | 306.96 | 322.94 |
| 2005 | 329.60 | 349.68 | 350.18 | 359.67 | 343.70 | 355.65 |
| 2006 | 370.98 | 370.70 | 371.16 | 390.40 | 380.00 | 386.83 |
| 2007 | 417.17 | 392.97 | 393.39 | 418.58 | 414.59 | 416.68 |
| 2008 | 427.05 | 416.58 | 416.96 | 444.57 | 446.37 | 445.21 |
| 2009 | 483.19 | 441.62 | 441.94 | 468.65 | 474.53 | 472.28 |
| 2010 | 509.04 | 468.15 | 468.41 | 491.03 | 498.69 | 497.54 |
| 2011 | 532.96 | 496.28 | 496.47 | 511.90 | 518.78 | 520.44 |
| 2012 | 568.99 | 526.11 | 526.21 | 531.40 | 535.07 | 540.13 |
| 2013 | 555.84 | 557.72 | 557.73 | 549.64 | 548.01 | 555.52 |
| 2014 | 548.41 | 591.23 | 591.14 | 566.73 | 558.09 | 565.28 |
| 2015 | 561.87 | 626.76 | 626.55 | 582.76 | 565.85 | 567.93 |
| MAPE(%) | | 8.09 | 8.11 | 4.61 | 3.81 | 3.66 |
| 2016 | 562.20 | 664.42 | 664.09 | 597.81 | 571.75 | 562.10 |
| 2017 | 550.38 | 704.35 | 703.87 | 611.94 | 576.20 | 546.77 |
| MAPE(%) | | 23.08 | 23.00 | 8.76 | 3.19 | 0.34 |

Table 2. The results by five grey models in forecasting total $CO_2$ emissions of Germany (unit: mt)

| Year | Data | GM | DGM | NGBM | FNGBM | CCFNGBM |
|---|---|---|---|---|---|---|
| 2008 | 806.5 | 806.50 | 806.50 | 806.50 | 806.50 | 806.50 |
| 2009 | 751.0 | 751.00 | 767.63 | 751.00 | 751.00 | 752.18 |
| 2010 | 780.6 | 780.60 | 767.03 | 780.60 | 780.60 | 758.17 |
| 2011 | 761.0 | 761.00 | 766.44 | 761.00 | 761.00 | 765.19 |
| 2012 | 770.3 | 770.30 | 765.85 | 770.30 | 770.30 | 770.13 |
| 2013 | 794.6 | 794.60 | 765.25 | 794.60 | 794.60 | 771.67 |
| 2014 | 748.4 | 748.40 | 764.66 | 748.40 | 748.40 | 770.16 |
| 2015 | 751.9 | 751.90 | 764.07 | 751.90 | 751.90 | 766.18 |
| 2016 | 766.6 | 766.60 | 763.48 | 766.60 | 766.60 | 760.22 |
| MAPE(%) |  | 1.64 | 1.64 | 1.51 | 1.36 | 1.52 |
| 2017 | 762.6 | 762.60 | 762.88 | 762.60 | 762.60 | 752.70 |
| 2018 | 725.7 | 725.70 | 762.29 | 725.70 | 725.70 | 743.96 |
| MAPE(%) |  | 2.54 | 2.56 | 2.19 | 3.39 | 1.91 |

Table 3. The results by nine comparative models in forecasting $CO_2$ emissions from fuel combustion of China (unit: mt)

| Year | Data | ARIMA(1,2,0) | PR(2) | ANN | SVM | GM | DGM | NGBM | FNGBM | CCFNGBM |
|---|---|---|---|---|---|---|---|---|---|---|
| 2001 | 3593.1 | 3591.49 | 3306.58 | 3451.82 | 4593.15 | 3593.10 | 3593.10 | 3593.10 | 3593.10 | 3593.10 |
| 2002 | 3910.6 | 3914.71 | 4083.19 | 3969.77 | 4988.34 | 5411.73 | 5422.82 | 3654.78 | 3910.51 | 3912.12 |
| 2003 | 4603.4 | 4236.88 | 4806.87 | 4571.96 | 5383.53 | 5654.53 | 5665.04 | 4603.98 | 4667.74 | 4680.63 |
| 2004 | 5413.4 | 5215.51 | 5477.62 | 5235.31 | 5778.71 | 5908.22 | 5918.07 | 5413.41 | 5412.98 | 5378.46 |
| 2005 | 6174.0 | 6198.20 | 6095.44 | 5923.98 | 6173.90 | 6173.29 | 6182.41 | 6109.25 | 6093.64 | 6007.43 |
| 2006 | 6757.2 | 6945.22 | 6660.33 | 6596.32 | 6569.09 | 6450.25 | 6458.56 | 6709.18 | 6698.80 | 6573.08 |
| 2007 | 7325.4 | 7378.54 | 7172.28 | 7214.46 | 6964.28 | 6739.63 | 6747.04 | 7226.33 | 7228.15 | 7080.43 |
| 2008 | 7457.4 | 7896.82 | 7631.31 | 7752.17 | 7359.46 | 7042.00 | 7048.41 | 7671.02 | 7685.18 | 7533.91 |
| 2009 | 7796.6 | 7683.18 | 8037.40 | 8197.86 | 7754.65 | 7357.94 | 7363.24 | 8051.68 | 8074.93 | 7937.51 |
| 2010 | 8231.7 | 8091.25 | 8390.57 | 8552.71 | 8149.84 | 7688.05 | 7692.12 | 8375.38 | 8403.00 | 8294.91 |
| 2011 | 8916.3 | 8646.18 | 8690.80 | 8826.30 | 8545.03 | 8032.97 | 8035.70 | 8648.18 | 8675.06 | 8609.54 |
| 2012 | 9090.0 | 9547.26 | 8938.10 | 9032.06 | 8940.21 | 8393.37 | 8394.63 | 8875.33 | 8896.69 | 8884.59 |
| 2013 | 9335.5 | 9373.54 | 9132.47 | 9183.93 | 9335.40 | 8769.93 | 8769.59 | 9061.44 | 9073.15 | 9123.09 |
| 2014 | 9329.6 | 9565.56 | 9273.91 | 9294.48 | 9730.59 | 9163.39 | 9161.30 | 9210.60 | 9209.38 | 9327.86 |
| 2015 | 9276.5 | 9377.75 | 9362.41 | 9374.14 | 10125.78 | 9574.50 | 9570.50 | 9326.45 | 9309.97 | 9501.56 |
| 2016 | 9230.3 | 9233.55 | 9397.99 | 9431.12 | 10520.96 | 10004.05 | 9997.98 | 9412.26 | 9379.09 | 9646.68 |
| 2017 | 9396.9 | 9182.62 | 9380.63 | 9471.67 | 10916.15 | 10452.88 | 10444.55 | 9470.96 | 9420.56 | 9765.55 |
| 2018 | 9606.6 | 9517.75 | 9310.35 | 9500.41 | 11311.34 | 10921.84 | 10911.07 | 9505.20 | 9437.86 | 9860.35 |
| 2019 | 9920.5 | 9648.43 | 9187.13 | 9520.74 | 11706.53 | 11411.85 | 11398.43 | 9517.40 | 9434.11 | 9933.12 |

Table 4. Errors (%) produced by nine comparative models in forecasting $CO_2$ emissions from fuel combustion of China.

| Year | ARIMA(1,2,0) | PR(2) | ANN | SVM | GM | DGM | NGBM | FNGBM | CCFNGBM |
|---|---|---|---|---|---|---|---|---|---|
| 2001 | 0.04 | 7.97 | 3.93 | 27.83 | 0.00 | 0.00 | 0.00 | 0.00 | 0.00 |
| 2002 | 0.11 | 4.41 | 1.51 | 27.56 | 38.39 | 38.67 | 6.54 | 0.00 | 0.04 |
| 2003 | 7.96 | 4.42 | 0.68 | 16.95 | 22.83 | 23.06 | 0.01 | 1.40 | 1.68 |
| 2004 | 3.66 | 1.19 | 3.29 | 6.75 | 9.14 | 9.32 | 0.00 | 0.01 | 0.65 |
| 2005 | 0.39 | 1.27 | 4.05 | 0.00 | 0.01 | 0.14 | 1.05 | 1.30 | 2.70 |
| 2006 | 2.78 | 1.43 | 2.38 | 2.78 | 4.54 | 4.42 | 0.71 | 0.86 | 2.72 |
| 2007 | 0.73 | 2.09 | 1.51 | 4.93 | 8.00 | 7.90 | 1.35 | 1.33 | 3.34 |
| 2008 | 5.89 | 2.33 | 3.95 | 1.31 | 5.57 | 5.48 | 2.86 | 3.05 | 1.03 |
| 2009 | 1.45 | 3.09 | 5.15 | 0.54 | 5.63 | 5.56 | 3.27 | 3.57 | 1.81 |
| 2010 | 1.71 | 1.93 | 3.90 | 0.99 | 6.60 | 6.55 | 1.75 | 2.08 | 0.77 |
| 2011 | 3.03 | 2.53 | 1.01 | 4.16 | 9.91 | 9.88 | 3.01 | 2.71 | 3.44 |
| 2012 | 5.03 | 1.67 | 0.64 | 1.65 | 7.66 | 7.65 | 2.36 | 2.13 | 2.26 |
| 2013 | 0.41 | 2.17 | 1.62 | 0.00 | 6.06 | 6.06 | 2.94 | 2.81 | 2.28 |
| 2014 | 2.53 | 0.60 | 0.38 | 4.30 | 1.78 | 1.80 | 1.28 | 1.29 | 0.02 |
| 2015 | 1.09 | 0.93 | 1.05 | 9.16 | 3.21 | 3.17 | 0.54 | 0.36 | 2.43 |
| 2016 | 0.04 | 1.82 | 2.18 | 13.98 | 8.38 | 8.32 | 1.97 | 1.61 | 4.51 |
| 2017 | 2.28 | 0.17 | 0.80 | 16.17 | 11.24 | 11.15 | 0.79 | 0.25 | 3.92 |
| MAPE(%) | 2.30 | 2.00 | 2.13 | 6.95 | 9.31 | 9.32 | 1.90 | 1.55 | 2.10 |
| 2018 | 0.92 | 2.96 | 1.06 | 17.05 | 13.15 | 13.04 | 1.01 | 1.69 | 2.54 |
| 2019 | 2.74 | 7.33 | 4.00 | 17.86 | 14.91 | 14.78 | 4.03 | 4.86 | 0.13 |
| MAPE(%) | 1.83 | 5.15 | 2.53 | 17.45 | 14.03 | 13.91 | 2.52 | 3.28 | 1.33 |

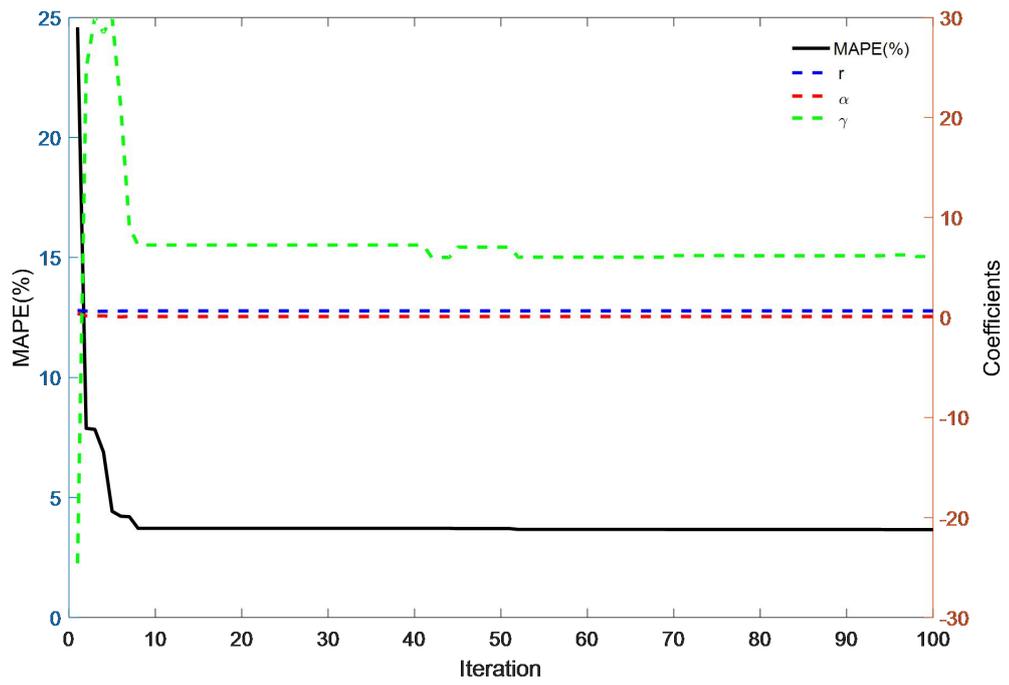

**Fig. 1**

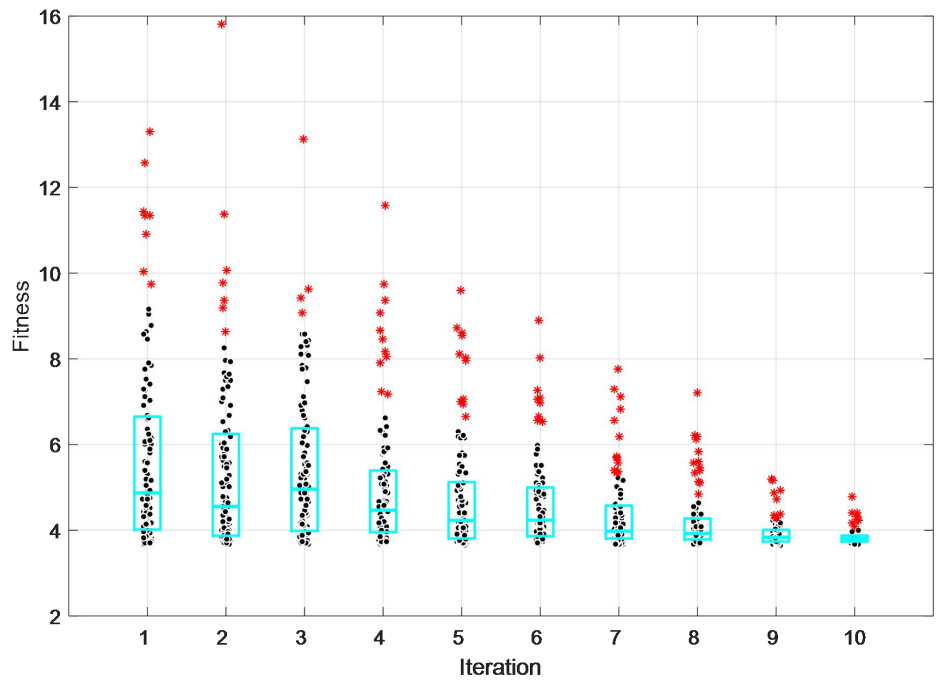

**Fig. 2**

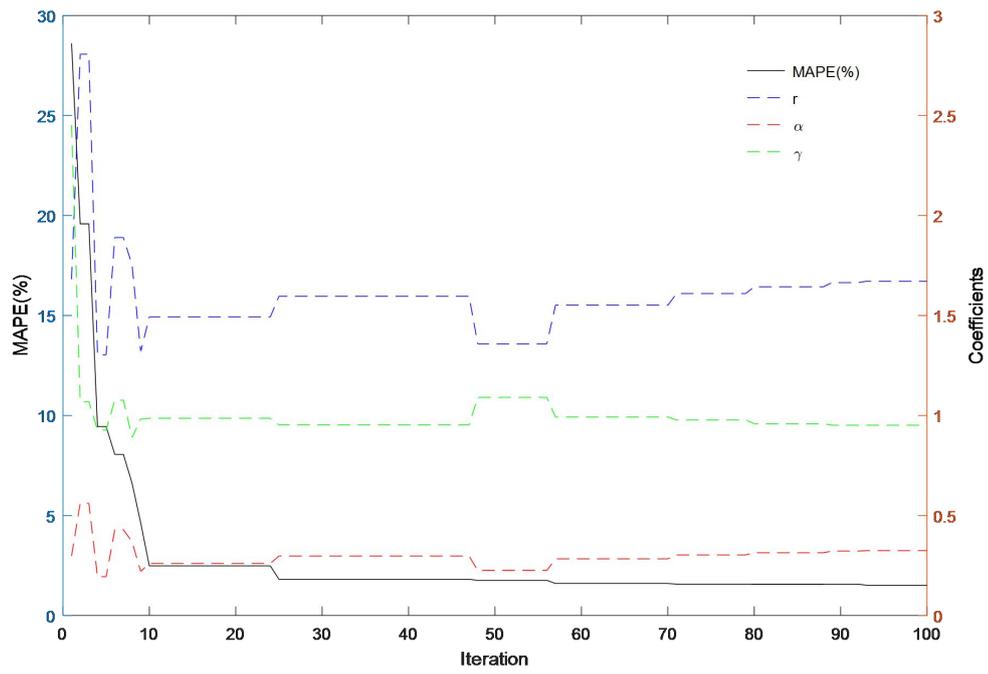

**Fig. 3**

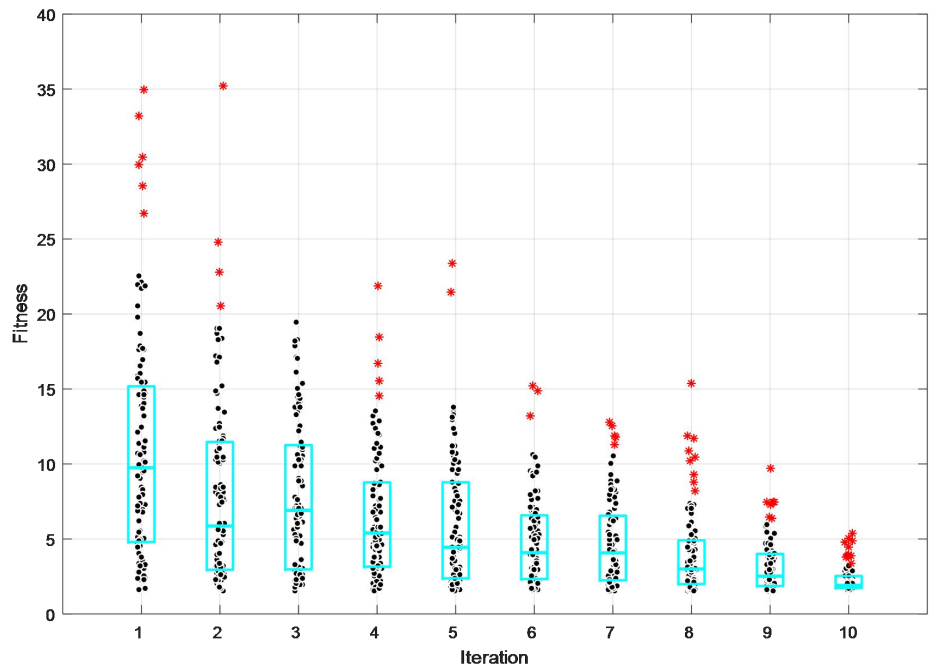

**Fig. 4**

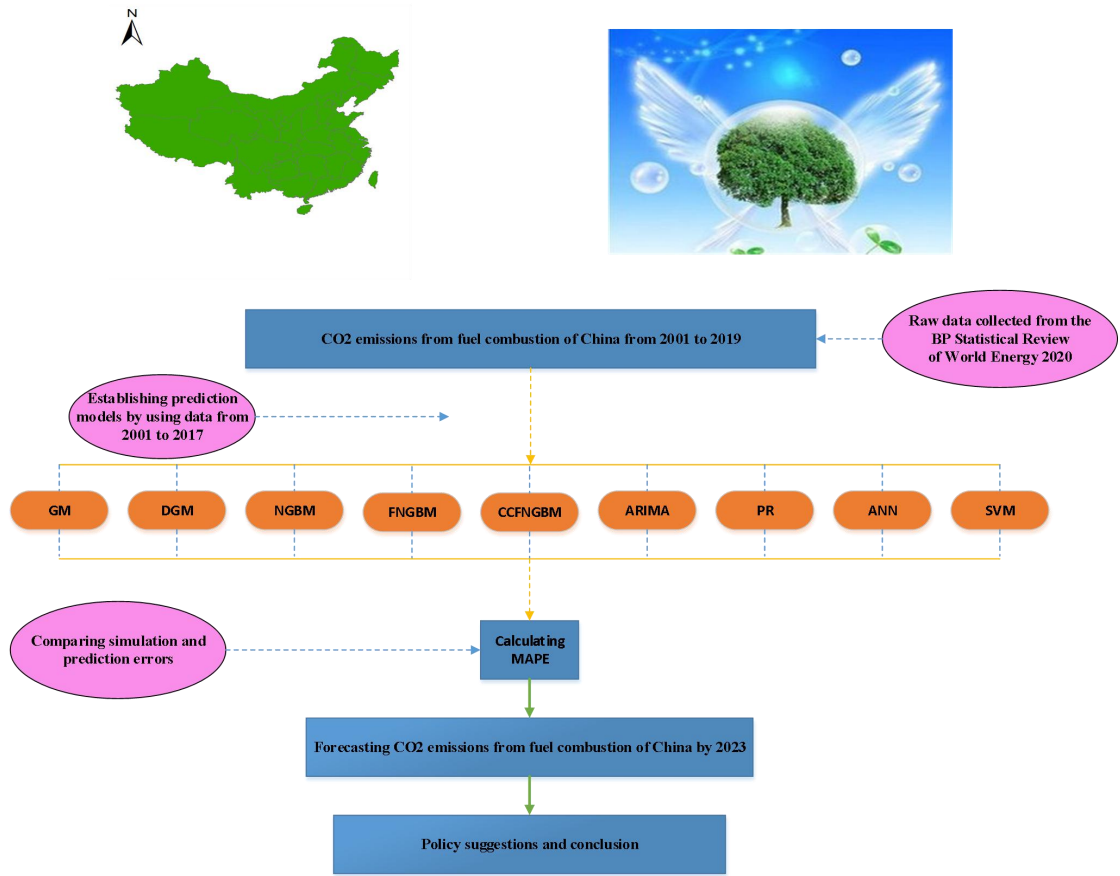

**Fig. 5**

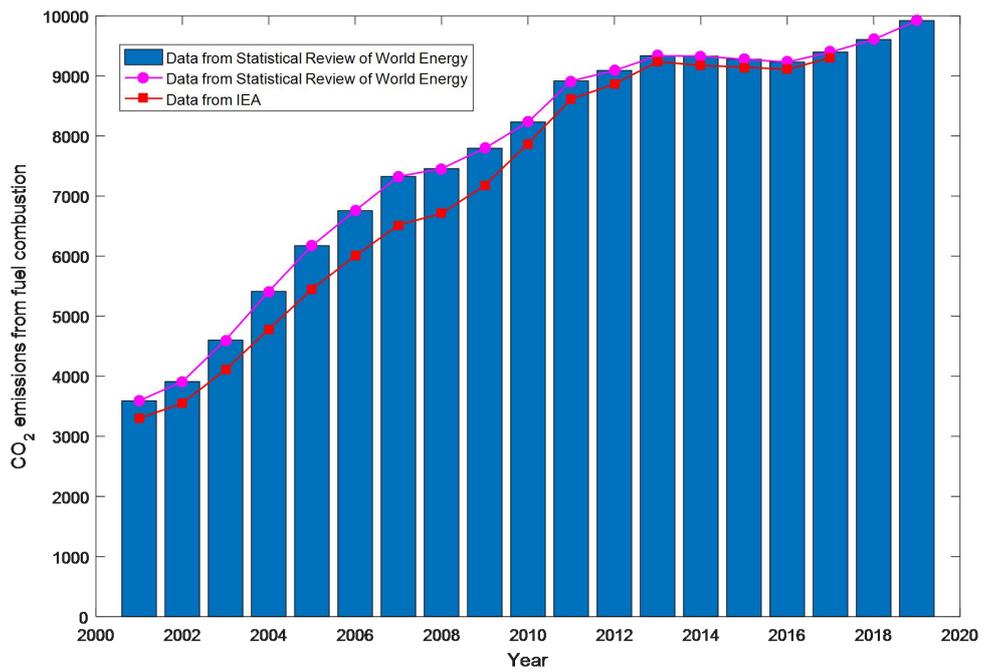

**Fig. 6**

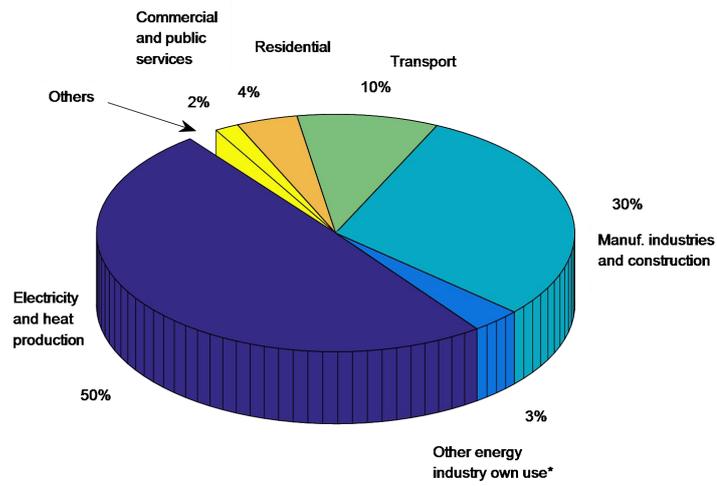

**Fig. 7**

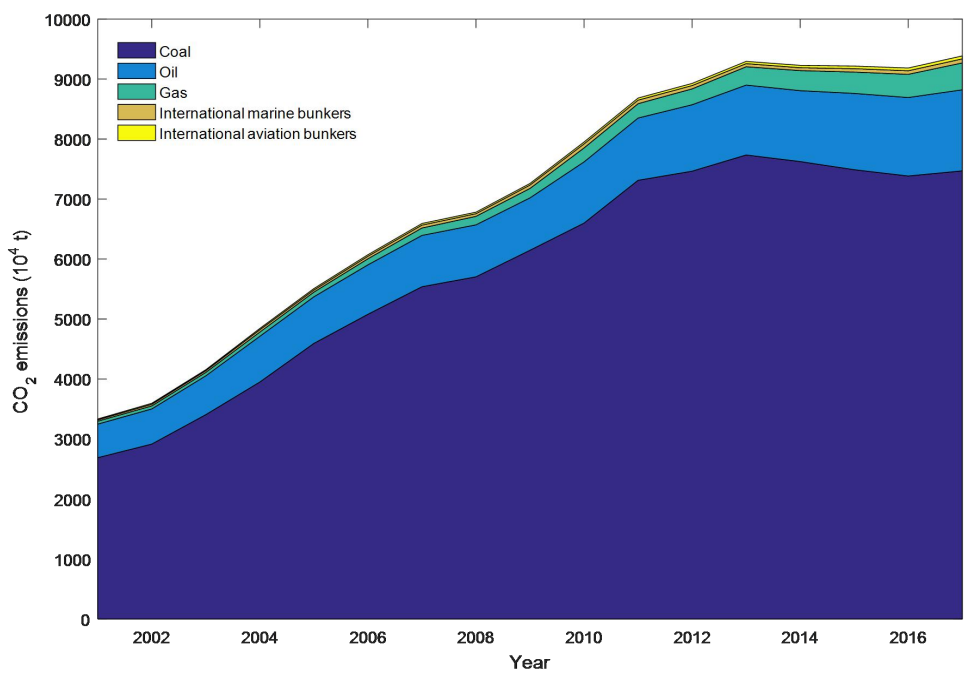

**Fig. 8**

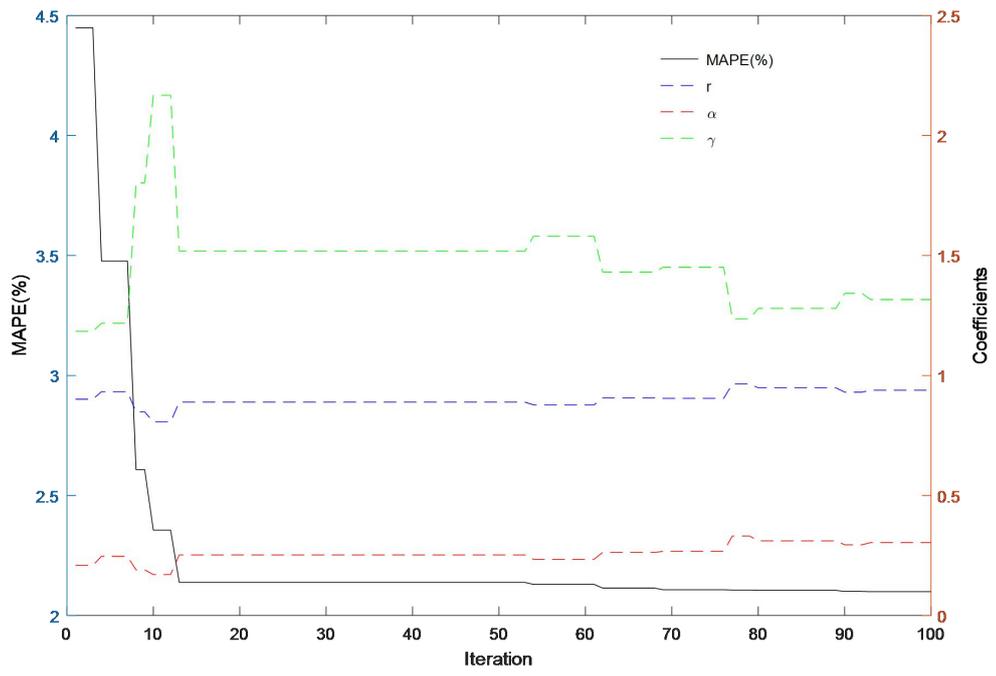

**Fig. 9**

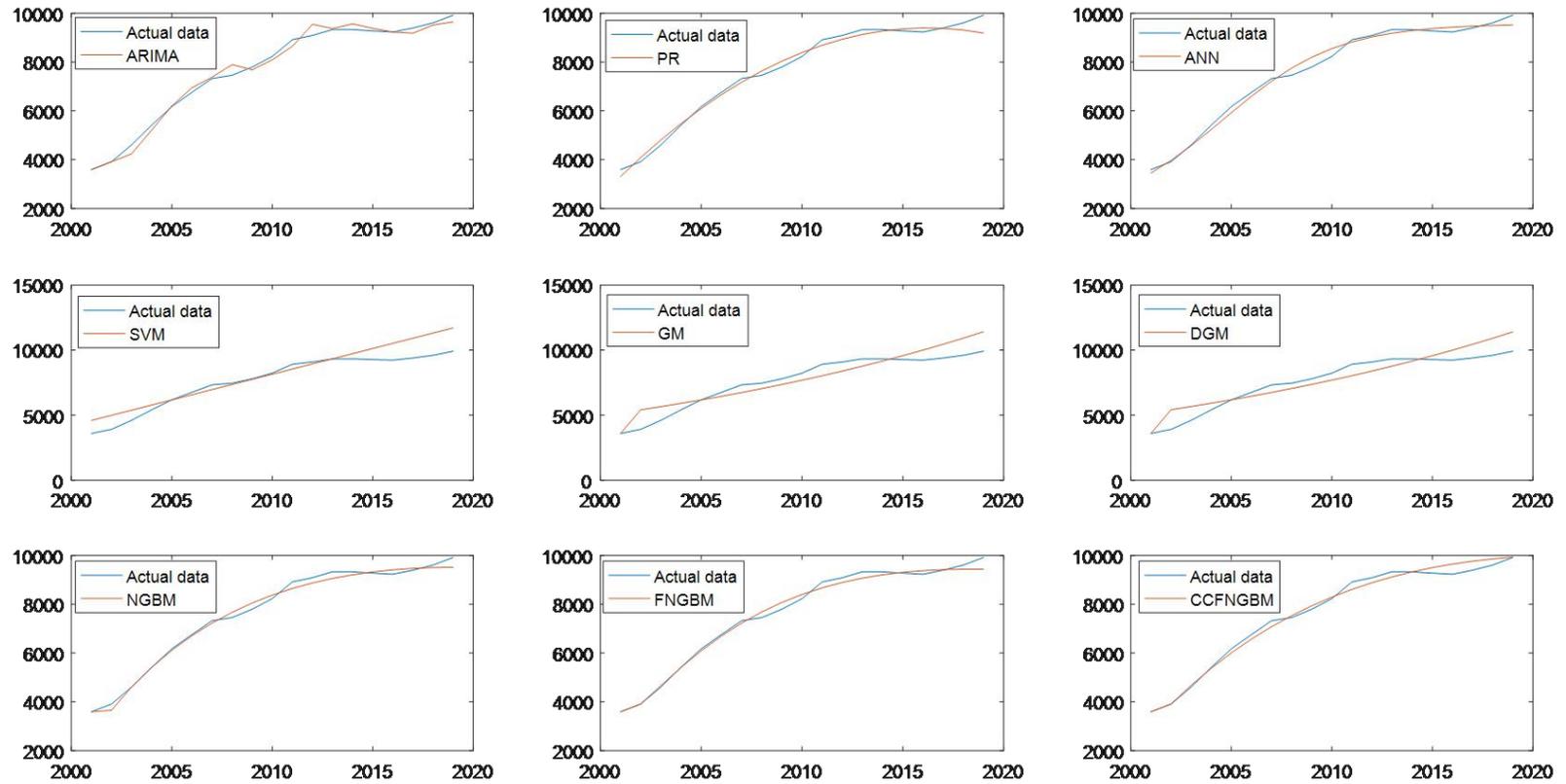

**Fig. 10**

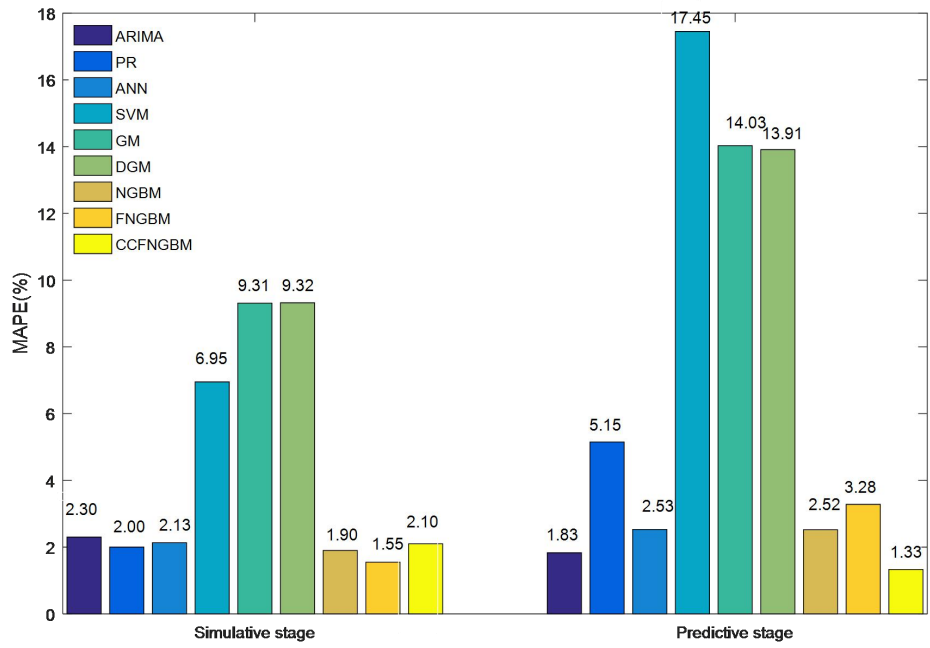

**Fig. 11**

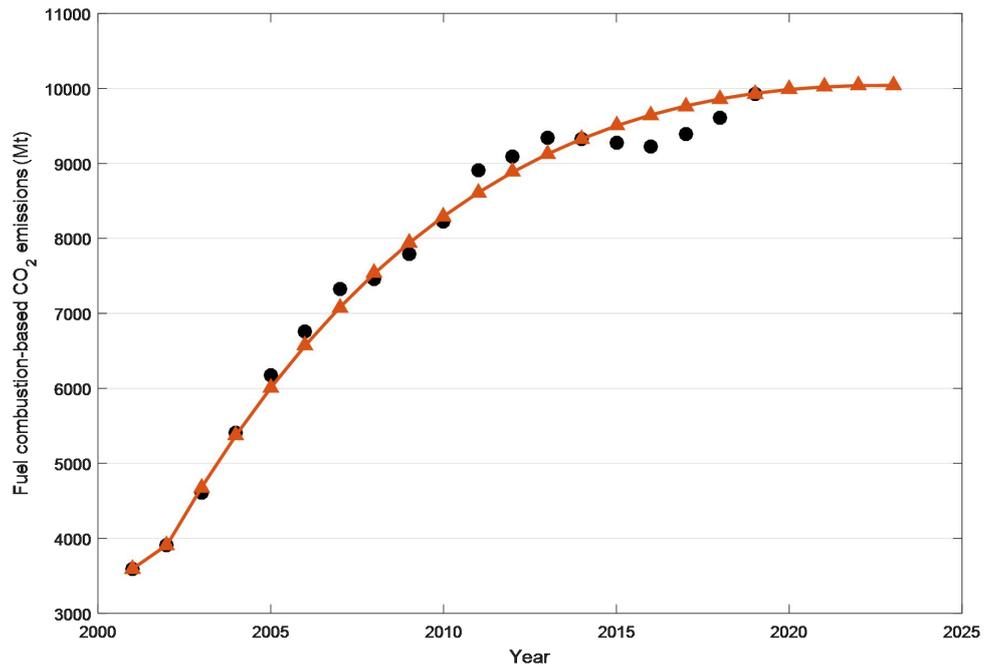

**Fig. 12**